\documentstyle[psfig]{mn}
\def\P3M{P$^3$M}
\def\vec#1{{\bf #1}}

\def\and{, }
\font\mgn=cmti7
\def\eqnote#1{\marginpar{\mgn #1}}
\def\eqnote#1{{}}

\newfam\cyrfam
\font\tencyr=wncyr10 at 12pt
\font\sevencyr=wncyr7 at 8.5pt \font\fivecyr=wncyr5 at 6pt

\textfont\cyrfam=\tencyr \scriptfont\cyrfam=\sevencyr
            \scriptscriptfont\cyrfam=\fivecyr

\def\fracnum#1#2{\raise 2.1pt\hbox{$\scriptstyle #1$}\kern
-1.2pt/\kern -1.2pt \lower 2.1pt\hbox{$\scriptstyle#2$}\,}

\begin{document}


\title[Cosmic shear] {The redshift and scale dependence of the cosmic
shear signal from numerical simulations} 

\author[Andrew J. Barber] {Andrew J. Barber$^1$\thanks{Email: 
abarber@pact.cpes.susx.ac.uk}\\
{}$^1$Astronomy Centre, University of Sussex, Falmer, Brighton, BN1 9QJ.
}

\date{Accepted 2001 ---. Received 2001 ---; in original form 2001 ---}

\maketitle

\begin{abstract}

The weak lensing shear signal has been measured numerically in
$N$-body simulations at 14 different redshifts ($z_s = 0.1$ to 3.6)
and on angular scales of $\theta = 2'$ to $32'$. In addition, the data
have been validated by analytical computations for an identical
cosmology, with density parameter $\Omega_m = 0.3$ and vacuum energy
density parameter $\lambda_0 = 0.7$. This paper reports on the scale
and redshift dependence of the shear variance, $\langle \gamma^2
\rangle$, which may be described by a simple formula of the form
$\langle\gamma^2\rangle(\theta,z_s) = a(\theta)z_s^{b(\theta)}$. The
redshift dependence for source redshifts up to 1.6, is found to be
close to $z_s^2$, which is a stronger dependence than earlier
analytical predictions ($\langle \gamma^2 \rangle \propto
z_s^{1.52}$), although, at higher redshifts, the $z_s$ dependence of
the shear variance is clearly less steep. The strong redshift
dependence further emphasises the need to know the precise redshift
distribution for the galaxy sources in any given survey, so that they
can be allocated to redshift bins accordingly, and the cosmic shear
signal correctly interpreted. Equations are also given for the
variance in the reduced shear, which is a more directly measurable
quantity observationally.

\end{abstract}


\section{INTRODUCTION}

Studies of weak gravitational lensing in cosmology are a very powerful
tool for attempts to understand the distribution of mass and the
evolution of the large-scale structure in the universe. Recently, it
has also become possible to constrain values for the cosmological
parameters from weak lensing studies. Since the gravitational
deflections of light arise from variations in the gravitational
potential along the light path, the deflections result from the
underlying distribution of mass which is usually considered to be in
the form of dark matter. The lensing signal therefore contains
information about the clustering of mass along the line-of-sight which
may be different from the clustering inferred from galaxy surveys
which trace the luminous matter only. In addition, by studying the way
light from large numbers of sources at high redshifts is deflected, it
is possible to obtain information about the way the clustering of mass
evolves with time.

As a result of weak gravitational lensing, a source at high redshift
will appear magnified (or de-magnified) as the beam converges (or
diverges) due to matter (or an under-density) contained within it. The
image also undergoes shearing due to deflections from matter outside
the beam, and causes a circular source, for example, to appear as an
elliptical image. Sources at similar redshifts and contained within a
small field-of-view will display similar magnification and shear
characteristics because their light will have passed along similar
density paths. For this reason there will be strong correlations in
the changes to the ellipticities, particularly on small scales, and
declining correlations on increasing angular scales.

The magnitude of this correlation depends strongly on the density
parameter and the value of the cosmological constant for the universe,
as these parameters reflect both the amount of mass and the rate of
evolution of structure. A number of attempts have been made to
estimate weak lensing statistics in different cosmological models,
both analytically and numerically. See Bernardeau, Van Waerbeke \&
Mellier (1997), Jain \& Seljak (1997) and Bacon, Refregier \& Ellis
(2000), for example, for analytical considerations, and Barber, Thomas
\& Couchman (1999), Jain, Seljak \& White (2000), Barber et
al. (2000), Hamana, Colombi \& Mellier (2000), Van Waerbeke et
al. (2001a) and Premadi et al. (2001), for example, for work done
using various cosmological $N$-body simulations. These authors have
attempted to predict values for various lensing statistics, many of
which may be obtained directly or indirectly from observational data.

A number of observational results have now been reported for the
so-called cosmic shear signal; see, for example, Bacon et al. (2000),
Kaiser, Wilson \& Luppino (2000), Maoli et al. (2001), Van Waerbeke et
al. (2000a, b), Wittman et al. (2000), Mellier et al. (2001), Rhodes,
Refregier \& Groth (2001) and Van Waerbeke et al. (2001b).  There is
also a number of ongoing and planned observational programmes such as
the Sloan Digital Sky Survey and surveys using the Visible and
Infra-red Survey Telescope for Astronomy (VISTA). These and other
telescope programmes will help to further constrain the cosmic shear
values and the cosmological parameters. An excellent review of weak
gravitational lensing and the measurement of the cosmic shear signal
in particular is contained in the paper by Bartelmann \& Schneider
(2001).

The purpose of this paper is to predict important cosmic shear
statistics using cosmological $N$-body simulations, which can be
compared directly with observations. In particular, it reports the
redshift and angular scale dependence for the shear variance and the
reduced shear, computed numerically and supported by analytical
computations. The detailed results are presented for background
sources at 14 different redshifts and angular scales from $2'$ to
$32'$. The finding that the shear variance has a strong redshift
dependence draws attention to the need to have good redshift data in
observational surveys before the cosmic shear signal can be correctly
interpreted in terms of the underlying cosmology.

The numerical method uses the algorithm described in Couchman, Barber
\& Thomas (1999), which computes the three-dimensional shear in the
simulations, and which has been applied to a simulation with density
parameter $\Omega_m = 0.3$ and vacuum energy density parameter
$\lambda_0 = 0.7$. (Cosmologies of this type will be referred to as
LCDM cosmologies.) To obtain the required statistics on different
angular scales, the computed shear values have been combined (using
the appropriate angular diameter distance factors and accounting for
multiple deflections) along lines of sight arranged radially from the
observer's position at redshift $z = 0$. To support the numerical
results, analytical computations have also been made for an identical
cosmology, using the procedure of Van Waerbeke et al. (2001a).
 
A brief outline of this paper is as follows.

In {\bf Section 2}, the equations and definitions required for weak
lensing are presented, and analytical work to determine weak lensing
statistics are described.

In {\bf Section 3}, the shear algorithm and the $N$-body simulations
are summarised together with details of the application of the code
for the generation of the lensing statistics at the observer's
location.

{\bf Section 4} presents the numerical results for the variances in
the shear and the reduced shear for sources at different redshifts,
and on different angular scales. Simple equations are presented which
fit the data to describe these variances as functions of redshift and
angular scale.

{\bf Section 5} is a discussion of the results.

\section{WEAK LENSING THEORY}

An explicit expression for each element of the three-dimensional shear
at an arbitrary position, $\vec{R}$, arising from matter at positions
$\vec{R}'$ has been given by Barber et al. (2000):
\begin{eqnarray}
\lefteqn{\frac{\partial^2\phi(\vec{R})}{\partial x_i\partial x_j} =
-\frac{4\pi G}{3} a^2 \bar{\rho} \,\delta_{ij}+
} \nonumber\\
\lefteqn{G\int\!\!\!\int\!\!\!\int\left[\frac{\rho(\vec{R}')}{\mid\vec{R}-\vec{R}'
\mid^3}\delta_{ij}
-\frac{3\rho(\vec{R}')(x_i-x_i')(x_j-x_j')}{\mid\vec{R}-\vec{R}'
\mid^5}\right]d^3R'}. \nonumber\\
 & & \hskip 2.8 in ~
\label{3dshear}
\end{eqnarray}
In this expression, $\phi$ is the peculiar gravitational potential,
$x_i$ and $x_j$ are position coordinates, with the suffixes $i$ and
$j$ representing the directions denoted by 1, 2 or 3, $G$ is the
universal gravitational constant, $\rho$ and $\bar{\rho}$ are the
density and mean density respectively, and $a$ is the scale factor for
the universe. The three-dimensional shear values computed by the shear
algorithm can be identified with this expression.

To compute the required physical properties at $z = 0$ arising from
the light of a distant source sheared by a {\em single} deflector, the
quantities needed are the two-dimensional second derivatives of the
effective lensing potential, which are denoted by $\psi_{ij}$, and
which are calculated by integrating the three-dimensional shear along
the coordinate direction $x_3$, and including the appropriate angular
diameter distance factors:
\begin{equation}
\psi_{ij} = \frac{D_d D_{ds}}{D_s}.\frac{2}{c^2} \int\frac{\partial^2
\phi(x_3)}{\partial x_i \partial x_j}dx_3.
\label{psiij}
\end{equation}
Here $D_d$, $D_{ds}$, and $D_s$ are the angular diameter distances
from the observer to the lens, the lens to the source, and the
observer to the source, respectively, and $c$ is the velocity of light
in vacuo.

Where there are multiple deflections, these quantities, evaluated for
the $i$th deflector ($i$ is now used as the deflector index), form the
elements of the shear tensor, ${\cal U}^{(i)}$, equivalent to the
derivatives of the reduced deflection angle,
\begin{equation}
{\cal U}^{(i)} = \left( \begin{array}{cc}
	\psi_{11}^{(i)}  & \psi_{12}^{(i)} \\
	\psi_{21}^{(i)}   & \psi_{22}^{(i)}
	\end{array}
	\right).
\label{U2}
\end{equation}
By combining the ${\cal U}^{(i)}$ for all the deflectors, the final
properties at $z = 0$ can be evaluated. To do this use is made of the
multiple lens-plane theory, which has been described concisely by
Schneider, Ehlers \& Falco (1992). In this theory, Jacobian
matrices, which describe how small changes in the position vector of
an element of the source relate to small changes in the position
vector of the corresponding element in the image, are constructed by
recursion. The final Jacobian matrix, $\cal A$, at $z = 0$ resulting
from $N$ deflectors is given by
\begin{equation}
{\cal A} = {\cal I} -\sum_{i=1}^{N} {\cal U} ^{(i)}{ \cal A} ^{(i)},
\label{Jacobian6}
\end{equation}
where $\cal I$ is the identity matrix, and in which the individual
Jacobian matrices are 
\begin{equation}
{\cal A} ^{(j)}={\cal I} -\sum _{i=1}^{j-1}\beta _{ij}{\cal U}^{(i)}
{\cal A}^{(i)}
\label{Jacobian7}
\end{equation}
for the $j$th lens, and
\begin{equation}
{\cal A} ^{(1)} = {\cal I}
\label{A1}
\end{equation}
for the first lens. In equation~\ref{Jacobian7},
\begin{equation}
\beta_{ij} \equiv \frac{D_s}{D_{is}}\frac{D_{ij}}{D_j},
\label{beta}
\end{equation}
where $D_s$ and $D_j$ are the angular diameter distances to the source
and the $j$th lens respectively, and $D_{is}$ and $D_{ij}$ are the
angular diameter distances from the $i$th lens to the source and the
$i$th lens to the $j$th lens respectively.  The final Jacobian can be
written in the form
\begin{equation}
{\cal A} = \left( \begin{array}{cc}
	1-\psi_{11}  & -\psi_{12} \\
	-\psi_{21}   & 1-\psi_{22}
	\end{array}
	\right),
\label{calA}
\end{equation}
from which the components of the overall two-dimensional shear,
$\gamma$, for weak lensing are
\begin{equation}
\gamma_1 = \frac{1}{2}(\psi_{11}-\psi_{22})
\label{gamma1}
\end{equation}
and
\begin{equation}
\gamma_2 = \psi_{21} = \psi_{12}.
\label{gamma2}
\end{equation}
(In a weak shear field smoothed by the variable particle softening and
where the gravitational potential and its derivatives are well-behaved
continuous functions, $\psi_{21} \simeq \psi_{12}$.)  The
two-dimensional shear is
\begin{equation}
\gamma = \gamma_1 + i\gamma_2
\end{equation}
($i \equiv \sqrt{-1}$), and the orientation of the major axis of the
resulting elliptical image is
\begin{equation}
\phi = \frac{1}{2}{\rm tan}\left(\frac{\gamma_2}{\gamma_1}\right).
\end{equation}
The effective convergence is
\begin{equation}
\kappa = \frac{1}{2}(\psi_{11}+\psi_{22}),
\label{kappa1}
\end{equation}
and the final magnification is
\begin{equation}
\mu =\left(\det \cal A \right)^{-1} = \frac{1}{(1-\kappa)^2-\gamma^2}.
\label{mu1}
\end{equation}

Since the shear causes the axes to be stretched by factors of
\begin{equation}
a = (1-\kappa - \mid \gamma \mid)^{-1}~~{\rm (major~~axis)}
\end{equation}
and
\begin{equation}
b = (1-\kappa + \mid \gamma \mid)^{-1}~~{\rm (minor~~axis)},
\end{equation}
the imposed ellipticity can easily be calculated from the elements of
the Jacobian matrix.  Observationally, the ellipticity of an image
is frequently defined in terms of the tensor of second brightness
moments,
\begin{equation}
Q_{ij}=\frac{\int {\rm d}\theta q_I
[I(\mbox{\boldmath$\theta$})](\theta_i - \bar{\theta_i})(\theta_j -
\bar{\theta_j})}{\int{\rm d}\theta q_I}[I(\mbox{\boldmath$\theta$})],
\end{equation}
where $I(\mbox{\boldmath$\theta$})$ is the surface brightness of the
galaxy image at angular position $\mbox{\boldmath$\theta$}$,
$\bar{\mbox{\boldmath$\theta$}}$ is the angular position of the centre
of light, and $q_I[I(\mbox{\boldmath$\theta$})]$ is a weighting
function in terms of the surface brightness. (See Blandford et
al., 1991, for example.) Then one definition is the complex ellipticity
\begin{equation}
\epsilon = \frac{Q_{11}-Q_{22}+2iQ_{12}}{Q_{11}+Q_{22}+2(Q_{11}Q_{22}-
Q_{12}^2)^{\frac{1}{2}}}.
\end{equation}
For elliptical isophotes, this definition is equivalent to
\begin{equation}
\epsilon = \frac{1-r}{1+r}{\rm e}^{2i\phi},
\end{equation}
where $r \equiv b/a$.

The ``reduced shear,'' $g$, evaluated for position
$\mbox{\boldmath$\theta$}$ in the image, is defined by
\begin{equation}
g(\mbox{\boldmath$\theta$}) \equiv
\frac{\gamma(\mbox{\boldmath$\theta$})}{1 -
\kappa(\mbox{\boldmath$\theta$})},
\label{g}
\end{equation}
so that the transformation between the source and image ellipticities
may be given by
\begin{equation}
\epsilon^{(s)} = \frac{\epsilon - g}{1-g^{\ast}\epsilon}
\end{equation}
for $\mid g \mid \leq 1$ (where the asterisk refers to the complex
conjugate).

Then, in the case of weak lensing, for which $\kappa$ and \mbox{$\mid \gamma
\mid \ll 1$}, \mbox{$\mid g \mid \ll 1$}, so that, for low intrinsic source
ellipticities, \mbox{$\epsilon \simeq \epsilon ^{(s)} + g$}.

However, the intrinsic ellipticities of observed galaxies are
generally unknown, so that the determination of the shear signal from
individual images is impossible. For this reason, it is necessary to
consider ensembles of galaxy images together, and to assume that the
galaxies of each ensemble have random intrinsic ellipticities, so that
the ensemble has zero net ellipticity.  Whilst this is strictly not
true, in high-redshift surveys in which the galaxies within each
narrow cone may be widely separated, it serves as a good working
approximation. A number of studies have been made into intrinsic
correlations of galaxy shapes; Heavens, Refregier \& Heymans (2000),
for example, have shown that the intrinsic correlation function for
elliptical galaxies at $z = 1$ in an LCDM cosmology is only of order
$10^{-4}$ on angular scales of $0.1'$ to $10'$, and is approximately
an order of magnitude lower than the correlations expected from weak
lensing. However, the intrinsic correlations are expected to exceed
those arising from weak shear in shallow surveys, as found, for
example by Brown et al. (2000), for sources with a median redshift of
only 0.1.

If then the ensemble of sources has zero net ellipticity, $\langle
\epsilon^{(s)}\rangle = 0$ and
\begin{equation}
\langle \epsilon \rangle = \frac{\sum_i u_i \epsilon _i}{\sum_i u_i}
\simeq g,
\end{equation}
where the $u_i$ are weight factors. Consequently then, in the case of weak
lensing only,
\begin{equation}
\gamma \simeq g \simeq \langle \epsilon \rangle ,
\end{equation}
and the variances in both the shear and the reduced shear for a
given angular scale are expected to be similar.

The importance of the convergence for understanding the evolution of
structure lies in its close association with the density contrast,
$\delta(\vec{x})$, at position $\vec{x}$, which is defined by
\begin{equation}
\delta(\vec{x}) \equiv \frac{\rho(\vec{x}) - \bar{\rho}}{\bar{\rho}}.
\end{equation}
By extension of the above equations for $\psi_{ij}$
(equation~\ref{psiij}) and $\kappa$ (equation~\ref{kappa1}), twice the
value of the effective convergence in the direction
$\mbox{\boldmath$\theta$}$ for a source at distance $x_s$ is
\begin{eqnarray}
2\kappa(\mbox{\boldmath$\theta$},x_s) & = &
\int_{0}^{x_s}\frac{D(x_3)D(x_s-x_3)}{D(x_s)}(\nabla^2-
\nabla_{x_s}^{2})\phi(\mbox{\boldmath$\theta$},x_3)
{\rm d}x_3 \nonumber\\
 & \simeq &
\int_{0}^{x_s}\frac{D(x_3)D(x_s-x_3)}{D(x_s)}\nabla^2\phi
(\mbox{\boldmath$\theta$},x_3){\rm d}x_3, 
\label{intdelta}
\end{eqnarray}
in which
\begin{equation}
\nabla^2\phi(\mbox{\boldmath$\theta$},x_3)=\frac{3H_0^2}{2}
\Omega_m\frac{\delta(\mbox{\boldmath$\theta$},x_3)}{a(x_3)},
\end{equation}
where $H_0$ is the Hubble parameter. Consequently, the effective
convergence represents a projection of the density contrast, and is
proportional to the density parameter, $\Omega_m$:
\begin{equation}
2\kappa(\mbox{\boldmath$\theta$},x_s) \simeq
\frac{3H_0^2}{2}\Omega_m\int_{0}^{x_s}\frac{D(x_3)D(x_s-x_3)}{D(x_s)}
\frac{\delta(\mbox{\boldmath$\theta$},x_3)}{a(x_3)}{\rm d}x_3.
\label{projection}
\end{equation}

It is also important to note how the two-point statistical properties
of the shear and convergence are related. From the
definitions of the individual shear tensors, ${\cal U}^{i}$
(equation~\ref{U2}), the components $\gamma_1$ and $\gamma_2$ of the
shear (equations~\ref{gamma1} and~\ref{gamma2}), and the effective
convergence, $\kappa$ (equation~\ref{kappa1}), the following
expressions apply in Fourier space:
\begin{equation}
\tilde{\gamma_1}(\vec{l}) = \frac{l_1^2-l_2^2}{l^2}\tilde{\kappa}(\vec{l}),
\end{equation}
and
\begin{equation}
\tilde{\gamma_2}(\vec{l}) = \frac{2l_1l_2}{l^2}\tilde{\kappa}(\vec{l}),
\end{equation}
where $l_1$ and $l_2$ are the components of the wavevector $\vec{l}$,
so that
\begin{equation}
\tilde{\gamma_1}^2(\vec{l}) + \tilde{\gamma_2}^2(\vec{l}) = 
\tilde{\kappa}^2(\vec{l}).
\end{equation}
Then it is clear that the power spectra for the shear,
$P_{\gamma}(l)$, and the convergence, $P_{\kappa}(l)$, are the same in
the case of weak lensing. 

To obtain values for the shear (or convergence) variances
analytically, the convergence power spectrum is integrated over all
wavenumbers, using a filter function appropriate for the required
angular scale, $\theta$. Since the convergence is obtained from a
projection of the density contrast (equation~\ref{projection}) from
the source redshift to the observer, the shear variance calculation
requires a complete spatial and temporal description of the matter
power spectrum, $P_{\delta}(k,x)$. This is a function of the
real-space wavenumber, $k~~(=l/(D(x)\theta)$, where $D(x)$ is the
angular diameter distance for a radial distance $x$ from the
observer. Kaiser (1998) has determined general
expressions for the angular power spectra of weak lensing distortions
for different cosmological models, and has estimated the growth of
this power with source redshift. As expected, the redshift dependence
is much stronger in low density cosmologies and especially so in
cosmologies dominated by a cosmological constant.

Jain \& Seljak (1997) give an expression equivalent to the following
for the shear variances derived analytically from the
matter power spectrum.
\begin{eqnarray}
\lefteqn{\langle \gamma^2 \rangle (\vartheta) =
\frac{1}{4}\times 36\pi^2\Omega_m^2\int_0^{\infty}k{\rm
d}k} \nonumber\\
 & & \hskip 0.3 in \times\int_0^{x_s}a^{-2}(x)P_{\delta}(k,x){\cal G}^2(x)W_2^2[kD(x)\vartheta]{\rm d}x,
\label{Jvar}
\end{eqnarray}
where $x_s$ is the radial distance to the source, 
\begin{equation}
{\cal G}(x)=\frac{D(x)D(x_s-x)}{D(x_s)}
\end{equation}
and 
\begin{equation}
W_2[kD(x)\vartheta] = 2J_1[kD(x)\vartheta]/[kD(x)\vartheta],
\end{equation}
where $J_1$ is the first Bessel function of the first order. The scale
$\vartheta$ is the angular radius of a circular window, so that the
formula has to be transformed to express the shear variance on scales,
$\theta$, represented by square pixels.

A complete description of the power spectrum on all scales (including
the linear, quasi-linear and non-linear regimes) is necessary for the
analytical approach, together with a detailed description of its
evolution. In particular, non-linear effects on scales of order $1'$
may increase the amplitude of the convergence power spectrum by an
order of magnitude. In addition, density fluctuations on scales
smaller than about $10'$ contribute most strongly to the weak lensing
signal, precisely where the non-linear evolution of the power spectrum
is most in evidence. In the non-linear regime, the fitting
formul\ae~of Peacock \& Dodds (1996), which extends the earlier work
of Hamilton et al., (1991) for the evolution of the matter correlation
function, may be used to map the non-linear wavenumbers onto
equivalent linear wavenumbers, and thus to evaluate the shear variance
values. These fitting formul\ae~include the stable-clustering
hypothesis which assumes an invariant mean particle separation on
sufficiently small scales. Analytical programs based on this
prescription are accurate to $\sim15\%$, depending on the cosmological
model.

For (circular) angular scales of $\vartheta = 2'$ and $15'$, Jain \&
Seljak (1997) summarised their findings in approximate power-law
expressions, which are equivalent to
\begin{equation}
\langle \gamma^2 \rangle[\vartheta = 2'(15')] \propto \vartheta^{-0.84}z_s^{1.52}\sigma_8^{2.58(2.00)}\Omega_m^{1.20(1.36)}
\label{Jain4}
\end{equation}
for LCDM cosmologies. The different indices for $\sigma_8$ and
$\Omega_m$ on the different scales enable the degeneracy between these
parameters to be lifted when measurements are made in both
regimes. The power of $z_s$ quoted is an intermediate value for the
two angular scales.

On the scales $1' < \vartheta < 30'$, the angular scale dependency was
similar, and the powers of $z_s$ were close; the index of $z_s$
decreased from 1.54 at $\vartheta = 2'$ to 1.48 at $\vartheta =
15'$. As a result, Jain \& Seljak (1997) found that, for LCDM
cosmologies,
\begin{equation}
\langle \gamma^2 \rangle(\theta) \propto \theta^{-0.84}z_s^{1.52}
~~~(1' < \theta < 30')
\end{equation}
approximately. ($\vartheta$ and $\theta$ are interchangeable in this
context.) 

The numerically determined shear variance results quoted in this
paper, and derived from the real-space values of the shear computed
numerically, will be compared with this approximate expression. In
addition, as a check of the validity of the numerical results,
analytical values for the shear variance have also been computed
directly for the precise cosmology used in the numerical
simulations. To achieve this the analytical program described in Van
Waerbeke et al. (2001a) has been used, which is a quite general program
for determining the weak lensing statistics in different
cosmologies. For the non-linear evolution of the power spectrum (and
the determination of variances), the fitting formul\ae~of Peacock \&
Dodds (1996) are used. For higher-order statistics, such as the weak
lensing skewness (not computed in this work), the code also computes
the evolution of the bispectrum at all scales, based on fitting
formul\ae~derived by Scoccimarro \& Couchman (2001) in numerical
simulations.

\section{NUMERICAL PROCEDURE}

Couchman et al. (1999) describe in detail the algorithm for the
computation of the elements of the matrix of second derivatives of the
gravitational potential in cosmological $N$-body simulations. It
computes all of the six independent component values of the
three-dimensional shear at each of the selected evaluation
positions. The rms errors in the computed shear component values are
typically $\sim 0.3\%.$

The algorithm uses a variable particle softening which distributes the
mass of each particle throughout a radius which depends on its
specific environment. The actual values of the softening parameters
are precisely as described in Barber et al. (2000), with the
minimum value set to $0.1h^{-1}$Mpc throughout, where $h$ is the
Hubble parameter expressed in units of 100km~s$^{-1}$Mpc$^{-1}$.

In the computation of the shear, the code uses the peculiar
gravitational potential, $\phi$, through the subtraction of a term
depending upon the mean density. This ensures that only light ray
deflections arising from departures from homogeneity apply, and is
equivalent to requiring that the net total mass in the system be set
to zero. The algorithm automatically includes the contributions of the
periodic images of the fundamental volume in computing the
three-dimensional shear at any location, thereby essentially creating
a realisation extending to infinity.

Since the algorithm works within three-dimensional simulation volumes,
rather than on planar projections of the particle distributions,
angular diameter distances to every evaluation position can be
applied. In this work it has been assumed that the angular diameter
distance varies linearly through the depth of each simulation volume.

The code has been applied to the cosmological $N$-body simulations of
the Hydra
Consortium\footnote{(http://hydra.mcmaster.ca/hydra/index.html)}
produced using the `Hydra' $N$-body hydrodynamics code (Couchman,
Thomas \& Pearce, 1995). Simulations of the LCDM Dark Matter only
cosmology were used with $\Omega_m = 0.3,$ $\lambda_0 = 0.7,$ power
spectrum shape parameter $\Gamma = 0.25$ and normalisation,
$\sigma_8$, on scales of $8h^{-1}$Mpc of 1.22. The number of
particles, each of mass $1.29 \times 10^{11}h^{-1}$ solar masses, was
$86^3$ and the simulation boxes had comoving side dimensions of
100$h^{-1}$Mpc. The simulation output times were chosen so that
consecutive simulation boxes could be abutted. A total of 48 boxes to
a redshift of 3.57 in the LCDM cosmology were used. To avoid obvious
structure correlations between adjacent boxes, each was arbitrarily
translated, rotated (by multiples of $90^{\circ}$) and reflected about
each coordinate axis, and in addition, each complete run was performed
20 times, so that averages of the final statistics were determined to
represent the required results.

To follow the behaviour of light rays from distant sources through the
simulation boxes, and obtain distributions of the properties at $z=0$,
a set of light paths was constructed emanating from the centre of the
front face of the $z = 0$ box and ending in a regular square array of
locations at the plane of the chosen source redshift.

A total of 14 source redshifts were selected to give good statistical
coverage of the redshifts of interest. These were redshifts of $z_s =
0.10,$ 0.21, 0.29, 0.41, 0.49, 0.58, 0.72, 0.82, 0.88, 0.99, 1.53,
1.97, 3.07 and 3.57. They corresponded to the simulation box redshifts
and were chosen to be close to redshifts of 0.1, 0.2, 0.3, 0.4, 0.5,
0.6, 0.7, 0.8, 0.9, 1.0, 1.5, 2.0, 3.0 and 3.5. In this paper the
source redshifts will be referred to loosely as the latter approximate
values, although in the determination of redshift dependences, etc.,
the actual redshift values were used.

Using a total of $317 \times 317$ lines of sight, the angular size of
the minimum particle softening is comparable to or less than the
angular separation of the adjacent lines of sight ($0'.49$) for all
redshifts greater than 0.14. Consequently, to account for the larger
angular size of the minimum softening at low redshifts, and also for
the allowed range of particle softening scales above the minimum
value, a resolution limit of $2'$ has been adopted for the data
analyses. In addition, the angular size of the gravitational force
softening used in the generation of the simulation boxes is below the
line-of-sight separation for all redshifts greater than 0.05. For a
source at redshift 1 in the LCDM cosmology, maximal gravitational
lensing occurs for a lens at redshift 0.36, and the angular separation
of adjacent lines of sight is approximately the same value as the
angular size of the minimum particle softening at that redshift
($0'.47$).

The total field-of-view for the set of lines of sight was $2.6^{\circ}
\times 2.6^{\circ}$, and this completely fills the near face of the
simulation box at redshift 1.0.

To establish the locations for the evaluation of the shear on each of
the lines of sight, first a regularly spaced (coarse) set of 50
locations was laid down on each line in each simulation box. Then
additional locations were computed at positions where the gravitational
potential was changing most rapidly, so that the potential field could
be well-sampled. To establish these locations, the particles were
assigned to volumes determined by a $10 \times 10$ grid within each
box. From each of the coarse line of sight locations, the separations
to the particles within the local grid volume and the nearest
neighbouring grid volumes was determined. If a separation was less
than the line-of-sight separation, a new evaluation location was
established on the line-of-sight, with coordinates corresponding to
the particle's position in the $x_3$ direction (radially from the
observer). All the evaluation locations (coarse and new) along each
line in each box were then sorted, labelled and counted so that the
programme to integrate the values along the lines of sight would
operate in the correct order for the correct number of locations.

Following the shear computations at all of the locations on all the
lines of sight in all the simulation volumes, the second derivatives
of the two-dimensional effective lensing potentials were obtained from
the three-dimensional shear values by integration, in accordance with
equation~\ref{psiij}. The integration was made in step-sizes
determined by the separation of adjacent evaluation locations on each
of the lines of sight, and so was different for every pair of
points. The integration algorithm was set to run from each of the
chosen source redshift planes, along each of the lines of sight to $z
= 0$, and values for the elements of the shear matrix, and thus the
Jacobian matrix, at the observer were obtained for each of the lines
of sight. From these data, all the required weak lensing statistics
were obtained for each line of sight and for each source redshift.

The full procedure, from the computation of the three-dimensional
shear values at all the evaluation locations to finally obtaining the
Jacobian matrices at $z = 0$, involved precisely the same
approximations as described fully in Barber et al. (2000). In
addition, it should be noted that, because the lines of sight project
radially from the observer at $z = 0$, some lines of sight pass
outside the confines of the simulation volume beyond a redshift of
1. In these cases, the periodicity of the boxes has been used to
reposition the lines of sight at equivalent locations within the
volumes.  The area of the far face of the most distant simulation box
(at redshift 3.6) is $1.3^{\circ} \times 1.3^{\circ}$. Consequently,
the procedure is not expected to introduce significant effects on the
computed variances on the scales of interest here, which are up to a
maximum of $32'$.

\section{RESULTS}

The individual real-space values for the shear, $\gamma$, and the
reduced shear, $g$, were computed from the final Jacobian matrices
obtained in each run for each line-of-sight and for each source
redshift. These data sets were then separately convolved with a
top-hat smoothing function of the required scale-size, and the
statistical variance values obtained on those specified scales. The
scale-sizes chosen for the top-hat smoothing were $2'$ (consistent
with the resolution limit of the numerical procedure),
$~~4',~~8',~~16'$ and $32'$. The computed values from each of the $N$
runs were then averaged, and the errors on the means of
$1\sigma/\sqrt{N}$ determined.

\begin{figure}
$$\vbox{
\psfig{figure=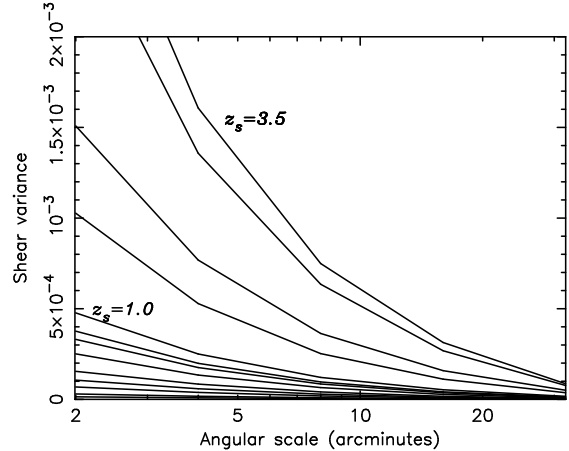,width=8.7truecm,angle=270}
}$$
\caption{
$\langle\gamma^2(\theta)\rangle$ for source redshifts 0.2 (lowest
curve), 0.3, 0.4, 0.5, 0.6, 0.7, 0.8, 0.9, 1.0, 1.5, 2.0, 3.0 and 3.5
(uppermost curve).
} 
\label{v3}
\end{figure}

Figure~\ref{v3} shows the shear variances computed in this way
(without the error bars for clarity) for all the source redshifts,
with the exception of $z_s = 0.1$, which is too low to be seen
clearly. The figure clearly emphasises the redshift dependence of the
results and suggests that a good knowledge of the redshift
distribution of the sources observed in surveys is essential to
interpreting the shear signal correctly. The relative closeness of
adjacent curves separated by intervals of only 0.1 in redshift also
suggests that the shear signal from sources with a {\em small} spread
in redshift {\em may} be adequately described by the shear signal
expected from sources at their median redshift. 

{\em However, it is not clear that the shear resulting from sources
with a significant redshift distribution will be representative of the
shear from sources at their median redshift.}




A number of recent measurements of cosmic shear have been made, as
mentioned in the Introduction. Following these measurements, Kaiser et
al. (2000) and Bartelmann \& Schneider (2001) have plotted them on a
single diagram and compared the results with the predicted values for
the shear variances as suggested by Jain \& Seljak (1997), whose work
is described above. As can be seen from the diagrams in these
references, the cosmic shear signal resulting from all the observed
measurements (from different telescopes, filters and cameras, and
different fields of view and data analysis techniques), appears to lie
very close to the formerly predicted values for sources at a redshift
of 1.

\begin{figure}
$$\vbox{
\psfig{figure=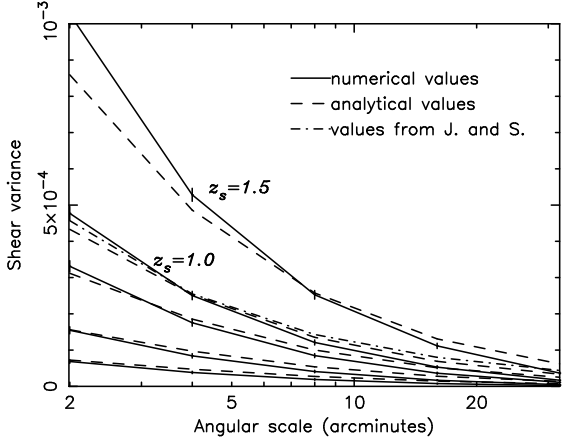,width=8.7truecm,angle=270}
}$$
\caption{ The numerical shear variances and analytical values computed
here, based on an identical cosmology and the same source
redshifts. Source redshifts of 0.4 (lowest pair of curves), 0.6, 0.8,
1.0 and 1.5 (uppermost pair of curves) are plotted, respectively. For
comparison, the analytical predictions of Jain \& Seljak (1997) for
sources at a redshift of 1 are also plotted, based on their
approximate general expression for LCDM cosmologies.  }
\label{varludojs_b}
\end{figure}

The numerical values computed here for source redshifts of 0.4, 0.6,
0.8, 1.0 and 1.5 are now plotted in figure~\ref{varludojs_b}, together
with the analytical values determined using Van Waerbeke et al.s
(2001a) code, described above; also shown are Jain \& Seljak's (1997)
predicted values (transformed to square pixel areas) for source
redshifts of 1, using their approximate expression for general LCDM
cosmologies. The advantage of Van Waerbeke et al's (2001a)
prescription is that the precise cosmological parameters and source
redshifts used in the numerical work have been used.

The numerically computed values for $z_s = 1$ are remarkably close to
the values predicted analytically by both Jain \& Seljak (1997) and
Van Waerbeke et al.s (2001a) prescription. In particular, the analytical
results computed here give strong support to the numerical results,
and the agreement is particularly good at low redshift and at
intermediate angular scales. The largest discrepancies occur only for
sources at high redshift (beyond about 1.5) and at angular scales
comparable with and below the resolution limit of the numerical data,
and where breakdown of both the numerical and analytical procedures
may also be expected. The values from Jain \& Seljak's (1997)
approximate general expression differ somewhat from both the numerical
and analytical results reported here for source redshifts other than 1,
so that it appears qualitatively that the results here would show a
different redshift dependence than the former analytical predictions.




\begin{figure}
$$\vbox{
\psfig{figure=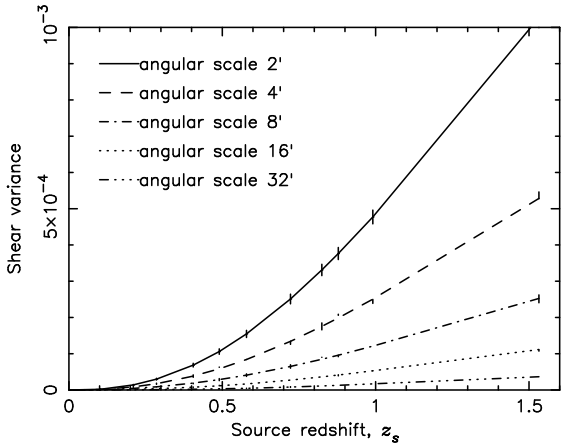,width=8.7truecm,angle=270}
}$$
\caption{
$\langle\gamma^2(\theta,z_s)\rangle$ as a function of source redshift for
angular scales of $2',~~4',~~8',~~16'$ and $32'$.
} 
\label{v4a}
\end{figure}

The redshift dependence of the shear variance on the chosen angular
scales is now plotted in figure~\ref{v4a}.




Since the redshift relationship for the shear variance and its scale
dependence are of fundamental importance to observational measurements
of the cosmic shear signal, the numerical results from
figures~\ref{v3},~\ref{varludojs_b} and~\ref{v4a} have been fitted
mathematically. It is assumed that the shear variance can be expressed
in the simple form
\begin{equation}
\langle\gamma^2\rangle(\theta,z_s) = a(\theta)z_s^{b(\theta)},
\label{gammathetaz}
\end{equation}
and indeed, for the redshift range $z_s \leq 1.6$, this form describes
the shear variances well. 

Over the range of angular scales $2'$ to $32'$, the coefficient
$a(\theta)$ can be expressed as
\begin{equation}
a(\theta) = (1.05\pm0.05)\times10^{-3}\theta^{-1.12\pm0.03},
\end{equation}
for $\theta$ in arcminutes. At the low end of the
range of scales, the index $b(\theta)$ falls very slightly as $\theta$ increases, and then remains almost constant
to beyond $32'$. Hence, for the entire range of scales $2'$ to $32'$,
$b(\theta)$ can be described well by the constant value
\begin{equation}
b(\theta) = 2.07 \pm 0.04.
\end{equation}









\begin{figure}
$$\vbox{
\psfig{figure=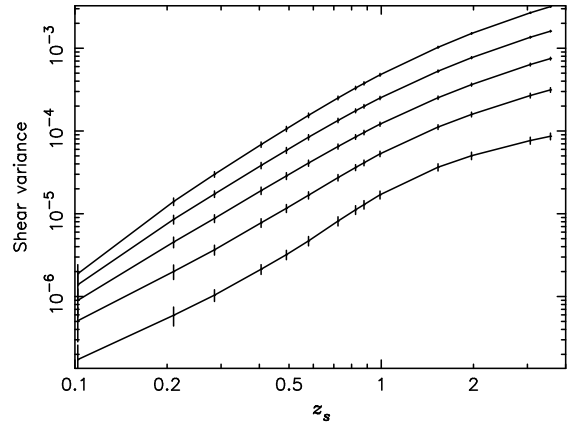,width=8.7truecm,angle=270}
}$$
\caption{
$\langle\gamma^2(\theta)\rangle$ vs. $z_s$ on logarithmic axes, to show the
gradually declining slopes at high redshift for the measured shear
variances. The curves are for angular scales of $2'$ (uppermost
curve), $4'$, $8'$, $16'$ and $32'$ (lowest curve).
} 
\label{v4alog}
\end{figure}

The relationships amongst $\langle \gamma^2 \rangle$, $\theta$ and
$z_s$ may be seen more clearly in figure~\ref{v4alog} for the
numerical data, which has logarithmic axes. At low redshifts
the slopes of the curves have a very small $\theta$ dependence, and,
for redshifts greater than about 1, the curves become clearly less steep,
indicating that the redshift dependence falls well below 2 at high
source redshifts.




In Section 2, it was shown that, in the case of weak lensing, both the
shear, $\gamma$, and the reduced shear, $g$, will be approximately
equal, so that the variances in both these quantities will also be
similar. Indeed, this approximation is used observationally to
estimate the values of the shear from the observed
ellipticities. However, the equality holds only in the weak lensing
limit. The numerical method used here for computing the weak lensing
statistics enables a direct comparison to be made between the
variances in $\gamma$ and $g$.

\begin{figure}
$$\vbox{
\psfig{figure=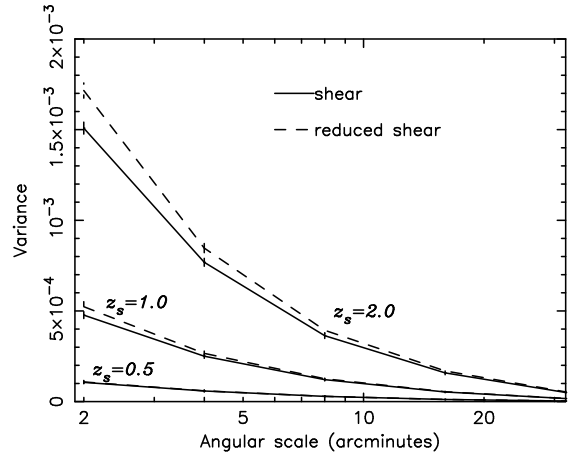,width=8.7truecm,angle=270}
}$$
\caption{
The variances in the reduced shear and the shear for comparison, for
source redshifts of 0.5, 1.0 and 2.0.
} 
\label{v2}
\end{figure}

Figure~\ref{v2} compares $\langle g^2\rangle(\theta)$ with $\langle
\gamma^2\rangle(\theta)$ for redshifts of 0.5, 1 and 2. For low source
redshifts, the curves are almost identical (except for the smallest of
angular scales). This is as expected, because this is the regime of
the weak lensing limit. However, departures between the two quantities
become increasingly obvious as the source redshift is increased.




\begin{figure}
$$\vbox{
\psfig{figure=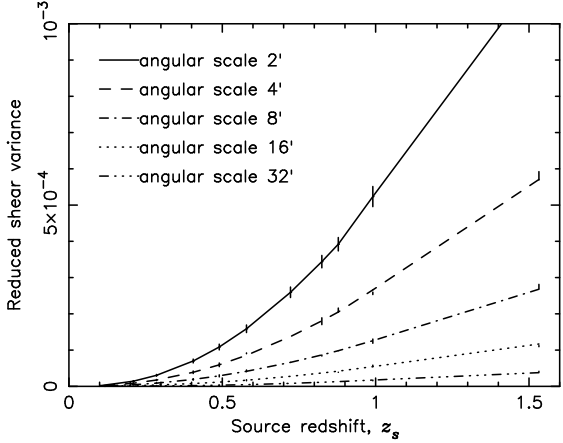,width=8.7truecm,angle=270}
}$$
\caption{
$\langle g^2(\theta,z_s)\rangle$ as a function of redshift for the
angular scales of $2'$, $4'$, $8'$, $16'$ and $32'$.
} 
\label{vea}
\end{figure}

The redshift dependence for $\langle g^2 \rangle$ on the different
angular scales is plotted in figure~\ref{vea}.




The functional form of $\langle g^2\rangle(\theta,z_s)$ can be written
in a similar way to that for $\langle\gamma^2\rangle(\theta,z_s)$. By
analogy with equation~\ref{gammathetaz}, the variance in the reduced
shear can be written as
\begin{equation}
\langle g^2\rangle(\theta,z_s) = c(\theta)z_s^{d(\theta)}
\label{gthetaz}
\end{equation}
in the range $z_s \leq 1.6$, for $2' \leq \theta \leq 32'$.

The coefficient $c(\theta)$ can be expressed as
\begin{equation}
c(\theta) = (1.2\pm0.1)\times10^{-3}\theta^{-1.21\pm0.03}.
\end{equation}
As for $b(\theta)$ in the expression for the shear variance,
$d(\theta)$ here falls very slightly as $\theta$ increases at small
angular scales, but
then remains almost constant throughout the range of angular
scales. Explicitly, for $2' \leq \theta \leq 32'$, $d(\theta)$ can be
expressed as the constant value
\begin{equation}
d(\theta) = 2.01\pm0.04.
\end{equation}

\begin{figure}
$$\vbox{
\psfig{figure=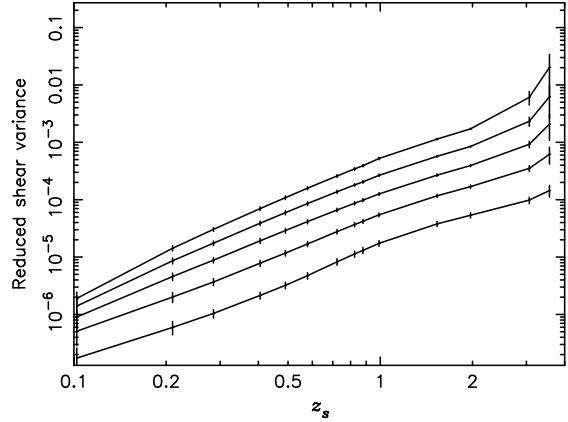,width=8.7truecm,angle=270}
}$$
\caption{
log($\langle g^2(\theta)\rangle$) vs. log($z_s$), to show the
gradually declining slopes throughout most of the redshift range for the reduced shear
variances on angular scales of $2'$ (uppermost curve), $4'$, $8'$,
$16'$ and $32'$ (lowest curve).
} 
\label{vealog}
\end{figure}

The variance in the reduced shear for the entire redshift range is
plotted with logarithmic axes in figure~\ref{vealog} 
to indicate both the $\theta$ and $z_s$ dependencies of the results.




As found for the shear variance, figure~\ref{vealog} for the reduced
shear, shows how the slopes of the
curves at the low redshift end have a slight $\theta$
dependence. However, unlike the shear, at high redshifts the slopes
become more noisy and generally steeper, even though the slopes do
decline very gently throughout the redshift range up to about 2. This
is most likely a result of the sensitivity to the convergence,
apparent from equation~\ref{g},
which becomes important either on small angular scales or at high
redshifts, where the lensing effects are strongest.

\section{DISCUSSION AND CONCLUSIONS}

The main results established in this paper are the dependences of the
shear variance and the variance in the reduced shear on source
redshift and angular scale. The numerical data obtained has been
validated by analytical computations using Van Waerbeke et al.'s
(2001a) program applied to an identical cosmology and the same source
redshifts.

For the shear variance, for source redshifts up to 1.6, one can write,
without the error bars for clarity,
\begin{equation}
\langle\gamma^2\rangle(\theta,z_s) = 1.05\times10^{-3}\theta^{-1.12}z_s^{2.07}
\end{equation}
for $2' \leq \theta \leq 32'$. I.e., the redshift dependence is close
to $z_s^2$, and stronger than the previous analytical predictions of
Jain \& Seljak (1997) who found a dependence close to
$z_s^{1.52}$. Whilst the above equation fits the numerical data well
within the range specified, the redshift dependence clearly begins to
decline for sources beyond a redshift of 1, so that the index of $z_s$
falls substantially below 2 at high redshift.

For the variance in the reduced shear, again it has been found that
the redshift dependence is close to $z_s^2$, and for $z_s \leq 1.6$ it
can be described by
\begin{equation}
\langle g^2\rangle(\theta,z_s) = 1.2\times10^{-3}\theta^{-1.21}z_s^{2.01}
\end{equation}
for $2' \leq \theta \leq 32'$, again without the error bars for
clarity. At higher redshifts, although there is an underlying tendency
for the curves to become less steep, the reduced shear is noisy and
sensitive to values of the convergence, so that the curves steepen
again at the high redshift end.

It was mentioned in the Results section that Kaiser et al. (2000) and
Bartelmann \& Schneider (2001) have plotted recent observational
determinations of the shear variances on a single diagram and compared
the results with the predicted analytical values of Jain \& Seljak
(1997). These plots show the observational results to lie close to the
predicted curve for sources at a redshift of 1. It should be noted at
this point that the values determined numerically here, and also
analytically using Van Waerbeke et al.'s (2001a) program, concur well
with the analytical values plotted {\em for source redshifts of 1}.
However, the observational results plotted have galaxies with
distributions in redshift; for example, Bacon et al. (2000) quote a
median redshift of $z_s = 0.8 \pm 0.2$ for their sample, Van Waerbeke
et al. (2000a) refer to a peak redshift of 0.9 for their galaxies,
Kaiser et al. (2000) has an ``effective'' redshift of 1.0, and the
survey of Maoli et al. (2001) has a broad redshift distribution which
peaks at a redshift of 0.8.

Whilst Jain et al. (2000) claim that a distribution of sources with a
mean redshift of 1 gives rise to an amplitude for the shear variance
only 10\% different from the case where all the galaxies are assumed
to lie at $z_s = 1$, the contention here is that it is necessary to
have a clear understanding of how the shear signal relates to the
redshift of the sources, and how a distribution of source redshifts
may further influence the conclusions. As an example, there are
differences of $\sim10\%$ -- 20\% between the shear variance values
for sources separated by a redshift interval of only 0.1 on angular
scales of $2'$. Attempts to constrain the cosmology precisely will
founder if the redshift dependence reported here is not taken into
account. According to equation~\ref{Jain4} an uncertainty of about
0.025 in the density parameter also results if there is a 10\%
uncertainty in the shear variance. In addition, there are clear
differences between the variances in the shear and the more directly
measured reduced shear, particularly at high redshift and on small
angular scales where the weak-lensing regime may not be
applicable. These factors point clearly to the need for studies into
the strength of the cosmic shear signal for sources distributed in
redshift, and will be investigated in a future paper.

The faster redshift dependence reported here ($\sim z_z^{2.07}$
compared with $\sim z_z^{1.52}$ previously predicted) is not easily
explained, as there are assumptions in both approaches. In the
numerical method one could point to the discontinuities in structure
as one passes from box to box, although the effect of this is reduced
statistically by employing a large number of runs (20 in this
work). The particle softening and resolution limitations might also be
a factor in the simulations. In the analytical approach, the main
difficulty arises in trying to describe accurately the evolution of
the power spectrum in the non-linear regime. Here mapping techniques
using fitting formul\ae, accurate to $\sim 15\%$, are used to relate
the linear and non-linear spectra, and assumptions such as the
stable-clustering hypothesis are used. Also, Jain \& Seljak's (1997)
published predictions are the result of summarising data from
different cosmological models, and then fitting the data to
approximate formul\ae~to express the power-law dependencies of the
statistics. Improvements have been made more recently however in
analytical procedures, for example, Seljak (2000) and Peacock \& Smith
(2000) have developed a model for the non-linear evolution of the
power spectrum based on the random distribution of dark matter haloes,
modulated by the large-scale matter distribution.

The main advantage of the numerical method is that the shear values
and variances are determined directly from the real-space data, rather
than from integration of a power spectrum which is less well
understood on the scales of interest. It is also expected that
numerical results may be used in the future to test the validity of
developments in analytical procedures, which will be increasingly
needed in view of the expected accuracies in the next generation of
cosmic shear measurements. For the interpretation of these
observations, fast analytical procedures will be more practical, since
it will be unrealistic to generate large numbers of numerical
simulations to cover a realistic parameter space.

A key influence on the redshift dependence of the shear variance lies
in the functional form of ${\cal G}(x)$, the angular diameter distance
function. ${\cal G}(x)$ occurs in equation~\ref{Jvar} where it
multiplies the matter power spectrum to produce the effective
convergence power spectrum. In the numerical method here it also plays
a crucial role in establishing the elements of the final Jacobian
matrices, from which the convergence and shear values are obtained. In
the numerical work here, the ``filled beam'' values for ${\cal G}(x)$
have been used as it is assumed that the light is always passing
through a region of smoothed density. (See Barber et al., 2000, for a
full discussion of angular diameter distances in inhomogeneous
universes.) The form of ${\cal G}(x)$ at high redshifts may also help
to explain the reducing index of $z_s$, because the redshift value at
the peak of the ${\cal G}(x)$ curve rises less quickly as the source
redshift is increased.

In this paper, no attempt has been made to simulate the signal which
might arise from typical noisy data. The intrinsic ellipticities of
the background galaxies is a source of random noise, and in addition,
measurement errors of very faint galaxies introduce errors in the
ellipticity estimates.

The intrinsic ellipticity correlations of galaxies in close proximity
to each other produce a real signal which has to be disentangled from
the cosmic shear correlations from the large-scale structure along the
line-of-sight. Heavens et al. (2000) have attempted to determine the
intrinsic correlations by modelling the shapes and angular momentum
vectors of galaxies in a way which reflects the shape and angular
momenta of their dark matter halos in $N$-body simulations. They find
that the intrinsic ellipticity correlation function for elliptical
galaxies is an order of magnitude below the expected weak lensing
signal on scales of $0.1'$ to $10'$ for sources at $z_s=1$ in a LCDM
cosmology. Brown et al. (2000) have analysed real data from the
SuperCOSMOS Sky Survey for galaxies with a low median redshift of $z_s
= 0.1$. They find that the ellipticity variance is approximately two
orders of magnitude higher than the expected weak lensing signal
throughout the range $1'$ to $100'$. As the source redshifts are low,
they claim that their result represents the real intrinsic
correlations since the weak lensing signal would be expected to be
small.

Taking the above two reports into account, there will be some
intermediate redshift at which the weak lensing signal overtakes the
intrinsic one. Consequently, whilst the result reported here presents
the pure redshift dependency of the weak lensing signal, it is clearly
advantageous to study good data at high redshift in order to
measure the uncontaminated cosmic signal.

This represents a further argument in favour of having good redshift
information in a galaxy survey designed to measure cosmic shear. It is
anticipated that forthcoming deep weak lensing surveys, such as that
proposed for the VISTA telescope, will also provide detailed
photometric redshift information to enable the surveyed galaxies to be
binned in redshift intervals. In this way it is anticipated that the
real cosmic shear signal may be interpreted more correctly.

\section*{ACKNOWLEDGEMENTS}

This work has been supported by PPARC and carried out with facilities
provided by the University of Sussex. The original code for the
three-dimensional shear computations was written by Hugh Couchman of
McMaster University. There were many useful discussions with Andy
Taylor, Antonio da Silva, Peter Thomas, Rachel Webster and Andrew
Melatos. The program to perfom the analytical computations was kindly
provided by Ludo Van Waerbeke.

\end{document}